# The GALAXIES Beamline at SOLEIL Synchrotron: Inelastic X-ray Scattering and Photoelectron Spectroscopy in the Hard X-ray Range.

J.-P. Rueff[a,b,c*], J. M. Ablett[a], D. Céolin[a], D. Prieur[a], Th. Moreno[a], V. Balédent[a†], B. Lassalle[a], J. E. Rault[a], M. Simon[a,b,c], and A. Shukla[c]

[a] *Synchrotron SOLEIL, L'Orme des Merisiers, 91192 Gif sur Yvette, France ;* [b]*CNRS, UMR 7614, Laboratoire de Chimie Physique-Matière et Rayonnement, F-75005, Paris, France ;* [c] *Sorbonne Universités, UPMC Univ Paris 06, UMR 7614, Laboratoire de Chimie Physique-Matière et Rayonnement, F-75005, Paris, France,* [d] *Institut de Minéralogie, de Physique des Matériaux et Cosmochimie, UMR CNRS 7590, UPMC Univ. Paris 06, Muséum National d'Histoire Naturelle, IRD UMR 206, 4 Place Jussieu, F-75005 Paris, France*

* Email : rueff@synchrotron-soleil.fr

† current address : *Laboratoire de Physique des Solides, CNRS-UMR 8502, Université Paris-Sud, 91405 Orsay, France*

**Synopsis**

The GALAXIES beamline at the SOLEIL synchrotron for inelastic x-ray scattering and photoelectron spectroscopy in the hard x-ray range.

**Abstract**

The GALAXIES beamline at the SOLEIL synchrotron is dedicated to inelastic x-ray scattering (IXS) and photoelectron spectroscopy (HAXPES) in the 2.3-12 keV hard x-ray range. These two techniques offer powerful, complementary methods of characterization of materials with bulk sensitivity, chemical and orbital selectivity, resonant enhancement and high resolving power. After a description of the beamline components and endstations, we address the beamline performances through a selection of recent works both in the solid and gas phases and using either IXS or HAXPES approaches. Prospects for studies on liquids are discussed.

**Keywords**
**Inelastic x-ray scattering ; RIXS ; Photoemission ; HAXPES**

**1. Introduction**

Inelastic X-ray Scattering (IXS) and HArd X-ray PhotoElectron Spectroscopy (HAXPES) have recently emerged as powerful methods for the characterization of materials. IXS and its resonant counterpart RIXS have followed the development of intense third generation synchrotron sources and high-resolution spectrometers. Similarly, HAXPES has become feasible with sufficient resolution only recently owing to the latest advances in photoelectron analyzers compatible with high kinetic energy. These two techniques are the main focus of the GALAXIES beamline at the SOLEIL synchrotron.

SOLEIL is a medium energy ring located in France on the Paris-Saclay scientific campus. The ring has been operational since 2008 and consists of 29 beamlines that operate in a broad range of wavelengths from THz radiation to hard x-rays. The GALAXIES beamline was selected in the second phase of construction to cover the field of IXS and HAXPES with an emphasis on solid-state physics but also encompassing gas phase studies and in the near future liquids. Each spectroscopic technique has its own endstation and dedicated sample environments. The beamline opened to users in May 2013.

Both HAXPES and RIXS provide detailed information about the electronic properties of materials, specific to the chemical element and selected orbitals. While RIXS is an all photon technique, well adapted to bulk samples and constrained sample environments (e.g. high pressure cells, furnaces and cryostats) because of the high penetration depth of x-ray photons, HAXPES retains the specificities and constraints of photoemission - surface sensitivity and the need for UHV conditions - but to a lesser extent as high energy electrons can escape from deep inside the solid (usually several tens of nanometers), and are thus less affected by the surface state. The research areas covered by the two endstations are widespread ranging from the study of correlated oxides, unconventional superconductors, geophysics, multilayers, interfaces and dilute phases.

**2. Beamline overview**

The GALAXIES beamline optical layout is designed to provide a highly tunable micro-focused x-ray source with high flux and high energy resolution in the 2.3-12 keV energy range (see Table 1 for technical details and Figure 1 for the schematic optical layout) on both the RIXS and HAXPES endstations. The beamline is located on a short straight-section equipped with an in-vacuum 20 mm period (98 periods) undulator. The U20 covers the energy range from 2.3 keV to above 12 keV using odd harmonics 1 through to 9 and the 2nd harmonic around 3.9 keV due to an energy "hole" between the 1st and 3rd harmonics (see Figure 2). The undulator magnet type was changed from $Sm_2Co_{17}$ permanent magnets ($B_{max}$=1.05 T) to higher field $Nd_2Fe_{14}B$ magnets ($B_{max}$=1.25 T) so as to reduce this energy "hole" width from 700 eV to 200 eV. The incident energy is selected by a Double Crystal Si111 liquid nitrogen cooled, fixed-exit Monochromator (DCM). The DCM is composed of a pseudo channel-cut arranged in the non-dispersive geometry. The vertical movement of the second crystal compensates for the beam height difference when scanning the energy, thus ensuring a fixed-exit beam.  An additional, 4-bounce, non-dispersive High-Resolution Monochromator (HRM) can be inserted into the beam to further narrow the energy bandwidth or higher order reflections of the DCM can equally be used. The HRM can host different crystals in order to function over the entire energy range while maintaining a resolution on the order of 100 meV. Details of the HRM are described elsewhere (Ablett et al., 2013). The collimating spherical mirror (M1) located upstream ensures maximum throughput through the HRM 4-bounce reflection. The mirror is mounted on a hexapod stage for accurate alignment in the beam. Carbon and palladium are available for the low energy (2.3 keV < E < 6 keV) and high energy (6 keV < E < 12 keV) spectral regions respectively which allows for the optimization of the mirror reflectivity and the elimination of high-order harmonics. Further downstream, the beam is focused onto the HAXPES station by the first Pd-coated toroidal focusing mirror (M2A) mounted on an identical hexapod stage. The M2A toroidal shape provides a spot of below 30 (V) x 80 (H) µm2 in size at the sample position in the HAXPES station. Unlike HAXPES, the RIXS endstation can be operated in either standard focusing mode using a similar optical scheme with a second Pd-coated toroidal mirror or micro-focusing mode with the help of vertical (VFM) and horizontal (HFM) focusing mirrors in the KB configuration while maintaining the sample focal plane with just a small difference in vertical height that is compensated for by moving vertically the whole RIXS spectrometer. The former mode can be selected by extracting the M2A mirror so as to send the beam directly onto the M2B mirror that focuses at the RIXS sample position. In the latter case, the M2A mirror is kept in place with the focused beam at the HAXPES station serving as a secondary source for the KB refocusing optics, and additionally inserting horizontally collimating slits to reduced to reduce by approximately three times the secondary horizontal source size. The KB optics are coated with both Rh and B4C layers and focus the beam to approximately 10 x 10 µm2 at the RIXS sample location. All of the mirror optics at GALAXIES have a fixed curvature and therefore provide excellent beam stability during experiments. Finally, after each optic, we have installed beamline diagnostics (x-ray imagers, quad beam-position monitors QBPM and photodiodes) in order to monitor the x-ray beam intensity and position. Each optic is equipped with piezo actuators (except the KB system) which we intend to implement in the near-future in a closed-loop feedback system in order to maintain the x-ray beam at the sample location within <5 microns over the whole energy range.

We are currently in the process of installing a diamond phase retarder immediately upstream of the DCM that will allow us to perform XLD (x-ray linear dichroism) and XMCD (x-ray magnetic circular dichroism) measurements (above 5keV) at both the HAXPES and RIXS end-stations.

**2.1 HAXPES endstation**

The HAXPES endstation is installed in the first experimental hutch. The performance and specifications of the endstation have been described in Céolin et al. (2013). The endstation is dedicated to hard x-ray, high kinetic energy photoelectron spectroscopy in the 0 – 12 keV kinetic energy region. The endstation consists of a main chamber under UHV conditions that hosts the electron analyzer, the sample manipulator and a gas cell (cf. Figure 3). Attached to the main chamber, the preparation chamber (also under UHV) serves for sample loading, basic sample treatment (heating, cleaving, sputtering, evaporation), surface characterization (LEED) and transfer to the main analysis chamber. The base pressure in both analysis and preparation chambers is $2 \times 10^{-9}$ mbar. The main chamber is mounted on a motorized frame for alignment in the x-ray beam. The photoemission spectra are recorded with a hemispherical SCIENTA EW4000 electron analyzer and a multi-channel plate coupled to a CCD detector. The analyzer is positioned in the horizontal plane with the lens axis parallel to the default polarization of the incident beam. The analyzer lens is specifically designed for high kinetic energy. The wide angular opening (45° or 60°) of the lens is the key feature of the analyzer. It significantly increases the number of electrons collected by the detector with respect to standard hemispherical analyzers, thus compensating for the weak cross-section at high kinetic energies while maintaining angular resolving capacity. Solid samples are mounted on Ta or Mo plates and are transferred to the main chamber by means of a linear feedthrough. The sample orientation in the main chamber can be controlled by a 4-axis, fully motorized manipulator. The manipulator is equipped with a He closed-cycle cryostat for low temperature measurements down to 15 K. The sample holder is electrically isolated from the manipulator for measurements of the sample current. Gases are introduced in the chamber through the gas cell, facing the sample manipulator. The typical gas pressure during measurements is of $10^{-5}$ mbar.

**2.2 RIXS endstation**

The RIXS endstation is installed in the second experimental hutch. It offers the possibility to perform IXS, RIXS or more simply X-ray emission spectroscopy (XES) experiments in the hard x-ray range and for a wide range of scattering angles in the vertical or horizontal planes thanks to its versatile design. Details of the endstation will be described elsewhere (Ablett et al., 2014). The endstation design is based on a 6-axis Newport diffractometer with two additional rotating arms – a long vertical arm and a short horizontal arm - for the energy and / or momentum analysis of the scattered photons (see Figure 3). The filtering of the emitted energy is realized by one or several spherically bent crystal analyzers, with bending radii of 0.5, 1 or 2m, arranged in the Rowland circle geometry and mounted on the spectrometer rotating arms. The horizontal arm can host a 4-analyzer assembly for IXS measurements while the vertical arm carries a single analyzer mainly for RIXS studies. The analyzers focus the scattered photons onto a detector mounted on a specific rotating arm. A He-filled bag is placed between the sample, the analyzers and the detector in order to reduce air absorption. The sample is positioned on a fully motorized 6-axis stage for precise alignment. Because of the large room around the sample position, the RIXS endstation can host a large variety of sample environments such as a goniometer head, a high pressure cell, a He cryostat, an oven, or a liquid cell – to cite some of the equipment that has already been utilized for RIXS measurements on the beamline For the detector, a large dynamic-range avalanche photodiode (APD) and a 2-dimensional detector

(MAXIPIX) are available. We will soon be implementing dispersion-corrected high-resolution IXS measurements with the MAXIPIX detector (Huotari et al., 2005). Normalization is obtained using either QBPM placed before the sample or using additional detectors (photodiodes or ionization chambers) measuring the scattered beam.

**3. Ancillary facilities**

The beamline instruments and data acquisition are controlled through a Python based command line interface running on top of the SOLEIL TANGO control system. This facilitates the writing of scripts for automatic multi-parameters acquisition. Data are available to the users either in either binary (NEXUS) or ASCII format and can also be downloaded from the SOLEIL User Network web-site after the experiment.

**4. Facility access**

The beamtime open access is granted through the standard proposal application of SOLEIL synchrotron or using private full price access. Details can be found on the SOLEIL User Network website (http://sunset.synchrotron-soleil.fr/sun/).

**5. Highlights**

The association of novel, high-resolution spectroscopic techniques with hard x-rays is the most salient feature of the GALAXIES beamline. We will focus on recent examples that illustrate this effective association.

**5.1 HARPES: Band structure in the bulk state**

The properties of materials are largely based on their electronic structure. The most popular method to visualize the electronic structure is certainly ARPES (angle-resolved photoemission spectroscopy). This allows the measurement of the kinetic energy of the photoelectrons and their momentum upon leaving the sample surface. When both kinds of information are combined, it is possible to reconstruct the electronic structure.

The development of ARPES was key to the study of fundamental physical phenomena such as electronic correlations, superconductivity and topological insulators. The measurements are traditionally performed with low energy photons in order to benefit form the high resolution in both energy and momentum space and a larger cross-section compared to hard-xrays. However, this requires perfectly clean surfaces, which limits the method to mostly two-dimensional or easily cleavable samples.

Recent studies have shown however that angular resolved photoemission is possible at higher kinetic energy, paving the way for the characterization of the bulk electronic structure of materials, a method referred to as HARPES (hard x-ray angle-resolved photoemission spectroscopy) (Gray et al., 2011) in reference to the ARPES technique. The first measurements on the GALAXIES beamline (see Figure 4) using the HAXPES endstation were obtained from a single crystal of diamond at 2.5 keV of kinetic energy at 20 K. Several dispersing bands can be clearly distinguished below the Fermi energy. This shows the feasibility of the method and opens up new possibilities beyond this demonstration experiment such as the study of the electronic structure of buried layers or interfaces.

**5.2 Raman scattering: Probing light elements in constrained environments**

K-edge XANES of light elements (such as B, C, O) is an effective probe of the chemical properties of the elements under study and their interaction with surrounding ligands or a solvent. However, their weak binding energies severely limits the experimental conditions that these edges can be measured with because of the necessity to use low energy x-rays. Inelastic x-ray scattering can overcome this difficulty. The K-edges are then observed in the energy loss domain like EELS (Electron Energy Loss Spectroscopy) but with the advantages of using hard x-rays (bulk sensitivity, large penetration depth), a method sometimes referred to as X-ray Raman Scattering (XRS) (Schülke, 2007).

Figure 5 illustrates this approach in $B_2O_3$. The measurements were carried out at 9721 eV incident energy using four Si(660) crystal analyzers arranged in a 2x2 matrix on the RIXS endstation. The intense elastic line at 0 eV energy loss is followed by the broad Compton peak, the B K-edge and the O K-edge of weaker intensity. Another advantage of XRS with respect to using soft x-ray XAS is the possibility to work beyond the dipole regime at high momentum transfers. This makes XRS sensitive to the orbital symmetry and orientation around the absorption site (Willers et al., 2012).

**5.3 Recoil effects in the gas phase**

In the case of an isolated atom, the photoionization at high photon energy induces a recoil of the ion by momentum conservation. This phenomenon is difficult to identify because of the large mass difference between the electron and the nucleus. To induce a strong effect, the photoelectron must have a high kinetic energy. At high energies, however, the spectral resolution decreases as well as the effective ionization cross sections, making it difficult to measure. An alternative is to measure the recoil effect not directly on the photoelectron spectrum, but on the Auger spectrum, which does not depend on the incident energy bandwidth. On the GALAXIES beamline, the high photon flux and large angular acceptance of the electron spectrometer can compensate the reduction in effective cross-section. These parameters have allowed us to make measurements of Auger spectra of ionized neon 1s shells and show that the recoil can be large enough for a Doppler splitting of the Auger lines to be possible (see Figure 6) (Simon et al., 2014). The results show a significant change of the Auger spectrum with the incident photon energy.

## 6. Discussion and conclusions

The GALAXIES beamline offers a unique combination of spectroscopic tools for the characterization of materials in the hard x-ray range.

New sample environments are under preparation to augment the beamline experimental capacity. The HAXPES endstation is soon to be equipped with a compact microjet setup for in-situ study on liquids. The microjet wil be mounted on the main chamber on the opposite flange to the analyzer. Beside the microjet itself, the setup consists of a beam catcher that traps the liquid jet in the chamber. Both the jet and the catcher are mounted on a manipulator for fine positioning. The liquid jet components can be removed for rapid servicing without the need to break the UHV in the main chamber. The instrument is expected be available in 2014. The setup will be adapted to the RIXS sample stage, opening new possibilities for the study of liquid phases. For solid samples, in-situ sample bias is readily available on the HAXPES station. The setup will be adapted to the RIXS station especially for the study of ferroelectrics materials.

**Acknowledgements**

We acknowledge the support groups of SOLEIL for their thorough assistance during the beamline installation, commissioning and operation. We are indebted to D. Chandesris for her initial impetus towards the development of HAXPES on the beamline and C. S. Fadley for his advice on how to push this further.

**Table 1** Beamline details

| Beamline name | GALAXIES |
|---|---|
| Source type | U20 in vacuum undulator, 20 mm magnetic period with 98 periods. Minimum gap = 5.5 mm; Kmax = 1.95 |
| Mirrors | Spherical collimating mirror (M1), R = 8.8 km; Si substrate with C and Pd coating<br>Toroidal focusing mirrors (M2A, M2B), R = 1.29 km, ρ = 30 mm; Si substrate with Pd coating<br>KB elliptical focusing mirrors: Vertical focusing mirror p = 5.9 m, q = 1.4 m, horizontal focusing mirror p = 6.3 m, q = 1.0 m; Si substrate with B4C or Rh coatings |
| Monochromator | Double crystal Si111 and 4-bounce High resolution monochromator (Si111 with an asymmetry angle (α) = 18°, Si220 α = 0° and 9°, and Si331, α = 0°) |
| Energy range (keV) | 2.3 – 12 keV |
| Wavelength range (Å) | 1.03 Å - 5.4 Å |
| Beam size (collimated, typical) (μm) | Standard focusing: 30 μm (V) x 80 μm (H); Micro focusing: 10 μm (V) x 10 μm (H) |
| Flux (standard focusing ) (photons s-1) | $3.0 \times 10^{14}$ ph/s/0.1%BW at 10 keV, 430 mA |
| Cryo capability | He cryostat 4K-300K (HAXPES / RIXS) |
| Sample mounting | HAXPES: 4 axis manipulator; RIXS: 6 axis sample stage |
| Detector type | HAXPES: Hemispherical analyzer EW4000. RIXS: Avalanche photodiode detector, silicon drift detector, 2D MAXIPIX. Diodes, QBPMs |
| RIXS 2θ capabilities | RIXS: 2θ ranging from 0° to 140° in horizontal plane and 0° to 130° in the vertical plane |

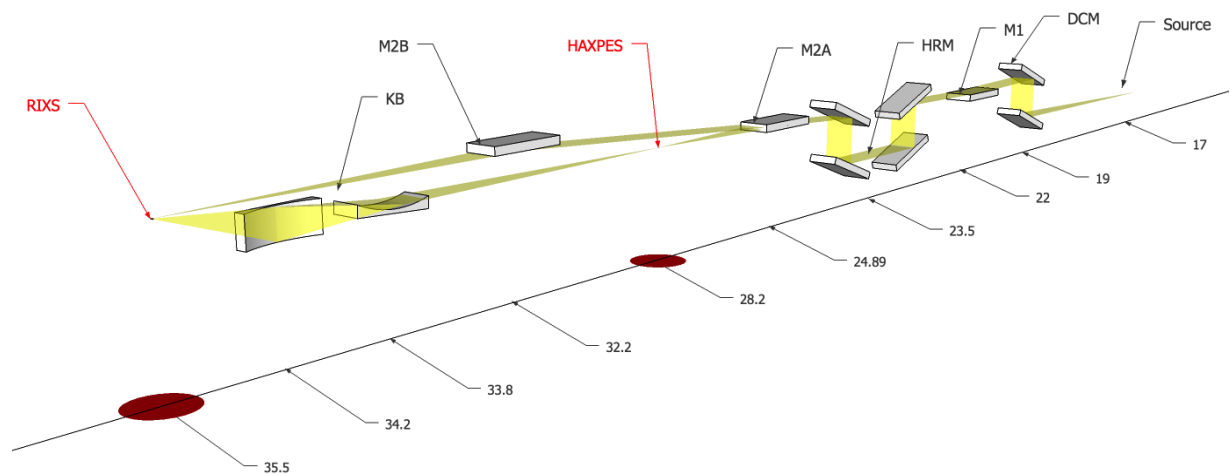

**Figure 1**  A schematic diagram of the beamline. Distance (in m) is taken from the center of the straight section.

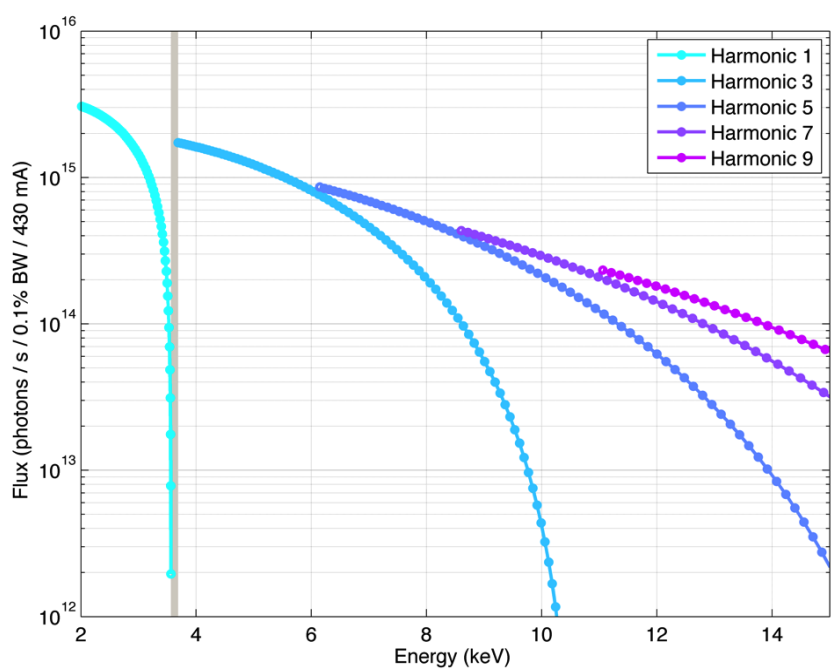

**Figure 2**  Flux simulations from the SPECTRA code (Tanaka & Kitamura, 2001) using a 1.8 mm (horizontal) x 0.6 mm (vertical) rectangular slit placed 11.7 m from the center of the U20, equivalent to the GALAXIES fixed-aperture diaphragm location in the beamline front-end. The shaded area indicates the energy 'hole' between the 1$^{st}$ and 3$^{rd}$ harmonics.

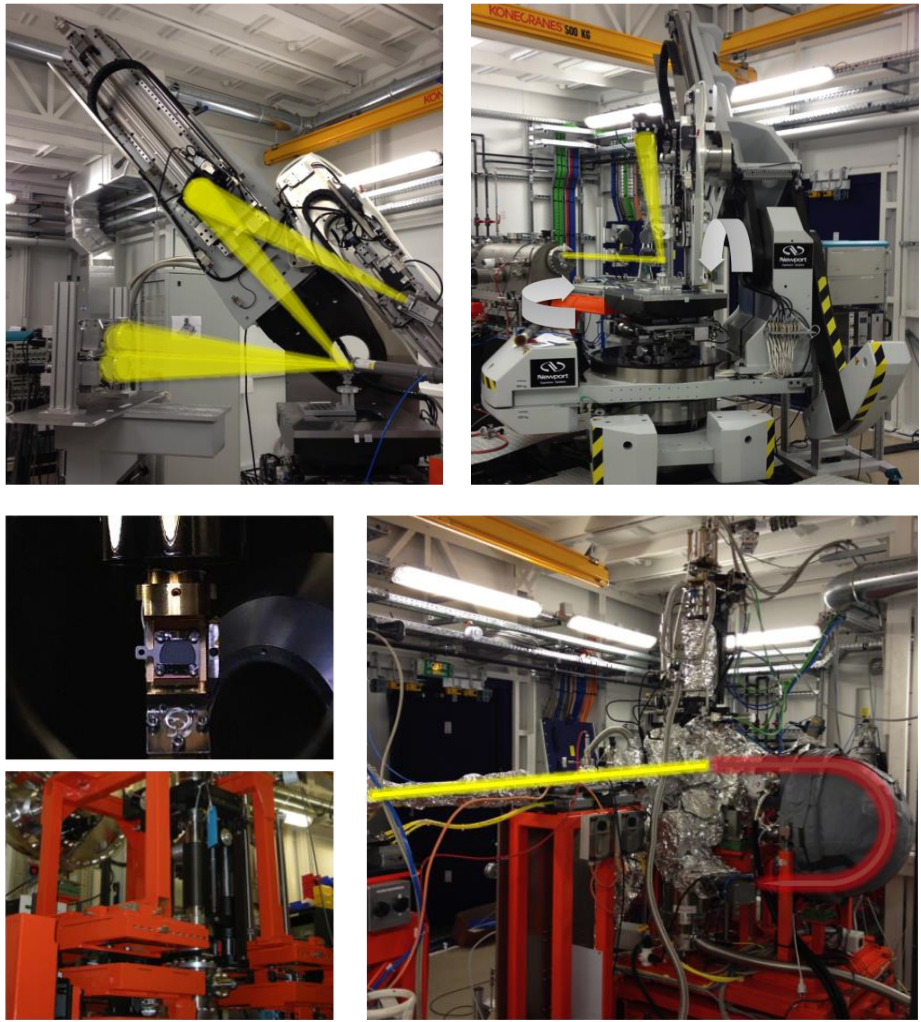

**Figure 3** (Top) Photos of the RIXS endstation with the multi-analyzer 2x2 setup and the long arm. The whole spectrometer can be rotated around the sample position – see arrows. (Bottom) Photos of the HAXPES endstation with the sample mounted on the manipulator, the gas cell and a view of the hemispherical analyzer.

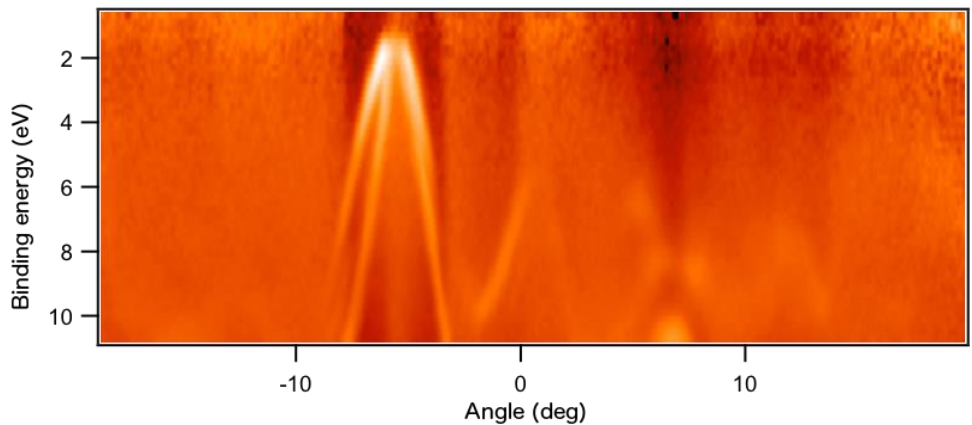

**Figure 4** High kinetic energy band structure or HARPES in doped diamond obtained on the GALAXIES beamline. The measurements were performed at 15 K with incident photons of 2.5 keV.

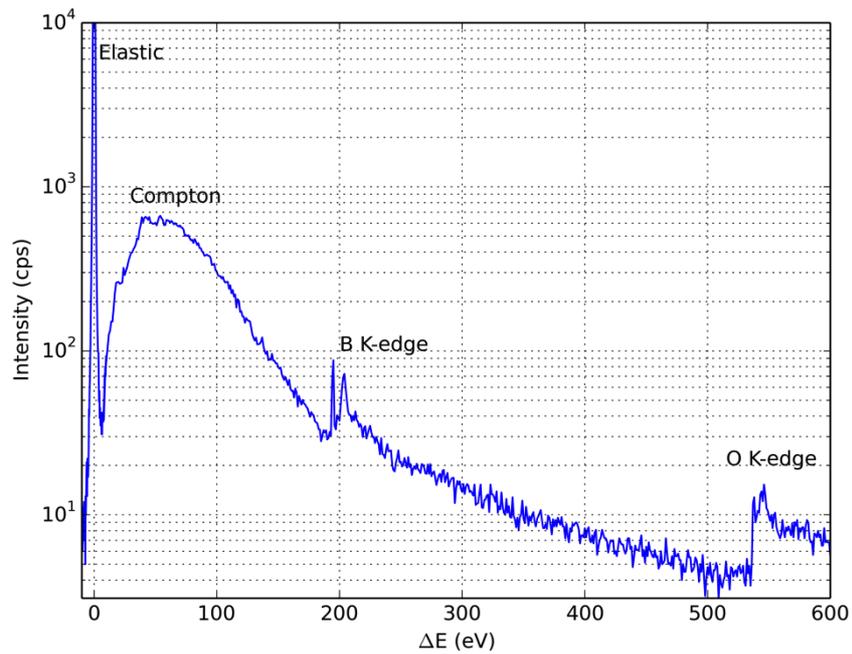

**Figure 5** X-ray Raman scattering spectra in B2O3 measured at 9721 eV as a function of the energy loss ΔE. Above the elastic line at 0 eV are the Compton peak and the B and O K-edges.

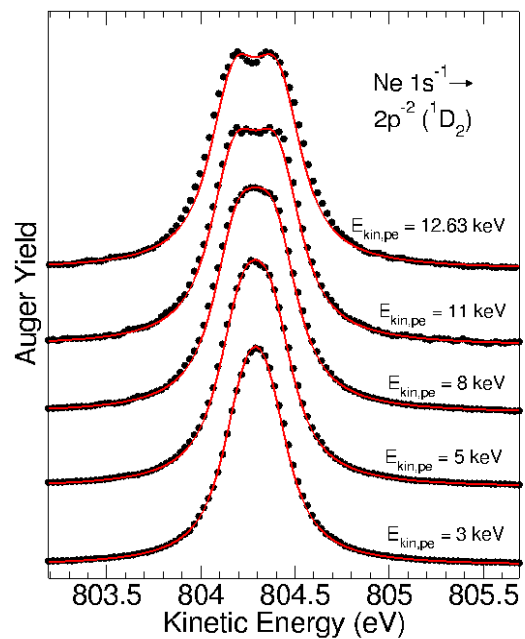

**Figure 6** Observation of recoil effect in Ne gas in the 1s-1 → 2p-2 Auger transition with increasing kinetic energies of the photoelectron Ekin,pe ranging from 3 keV to 12.63 keV: Experimental results (dots) and numerical simulations (red) (from Simon et al, 2014)